\documentclass{iaus}
\usepackage{graphicx}

\title[Star cluster evolution in the Magellanic Clouds revisited]
{Star cluster evolution in the Magellanic Clouds revisited}

\author[Richard de Grijs \& Simon P. Goodwin]{Richard de Grijs$^{1,2}$
\and Simon P. Goodwin$^1$}

\affiliation{$^1$Department of Physics \& Astronomy, The University of
Sheffield, Hicks Building, Hounsfield Road, Sheffield S3 7RH, UK \\
email: {\tt [R.deGrijs,S.Goodwin]@sheffield.ac.uk} \\[\affilskip]
$^2$National Astronomical Observatories, Chinese Academy of Sciences,
20A Datun Road, Chaoyang District, Beijing 100012, China}

\pubyear{2008}
\volume{256}  
\pagerange{1--6}
\jname{The Magellanic System: Stars, Gas, and Galaxies}
\editors{Jacco Th. van Loon \& Joana M. Oliveira, eds.}
\begin{document}

\maketitle

\begin{abstract}
The evolution of star clusters in the Magellanic Clouds has been the
subject of significant recent controversy, particularly regarding the
importance and length of the earliest, largely mass-independent
disruption phase (referred to as `infant mortality'). Here, we take a
fresh approach to the problem, using a large, independent, and
homogeneous data set of $UBVR$ imaging observations, from which we
obtain the cluster age and mass distributions in both the Large and
Small Magelanic Clouds (LMC, SMC) in a self-consistent manner. We
conclude that the (optically selected) SMC star cluster population has
undergone at most $\sim$30\% (1$\sigma$) infant mortality between the
age range from about 3--10 Myr, to that of approximately 40--160 Myr.
We rule out a 90\% cluster mortality rate per decade of age (for the
full age range up to $10^9$ yr) at a $>$6$\sigma$ level. Using a
simple approach, we derive a `characteristic' cluster disruption
time-scale for the cluster population in the LMC that implies that we
are observing the {\it initial} cluster mass function
(CMF). Preliminary results suggest that the LMC cluster population may
be affected by $<10$\% infant mortality.
\end{abstract}

\firstsection 

\section{Introduction}

One of the most important diagnostics used to infer the formation
history, and to follow the evolution of an entire star cluster
population is the `cluster mass function' (CMF; i.e., the number of
clusters per constant logarithmic cluster mass interval, ${\rm
d}N/{\rm d}\log m_{\rm cl}$). The {\it initial} cluster mass function
(ICMF) is of particular importance. The debate regarding the shape of
the ICMF, and of the CMF in general, is presently very much alive,
both observationally and theoretically. This is so because it bears on
the very essence of the star-forming processes, as well as on the
formation, assembly history, and evolution of the clusters' host
galaxies on cosmological time-scales. Yet, the observable at hand is
the cluster {\it luminosity} function (CLF; i.e., the number of
objects per unit magnitude, ${\rm d}N/{\rm d}M_V$).

The discovery of star clusters with the high luminosities and the
compact sizes expected for (old) globular clusters (GCs) at young ages
facilitated by the {\sl Hubble Space Telescope (HST)} has prompted
renewed interest in the evolution of the CLF (and CMF) of massive star
clusters. Starting with the seminal work by Elson \& Fall (1985) on
the young Large Magellanic Cloud (LMC) cluster system (with ages
$\lesssim 2 \times 10^9$ yr), an ever increasing body of evidence
seems to imply that the CLFs of young massive clusters (YMCs) are well
described by a power law of the form ${\rm d}N \propto L^{1+{\alpha}}
{\rm d}\log L$, equivalent to a cluster luminosity spectrum ${\rm d}N
\propto L^{\alpha} {\rm d} L$, with a spectral index $-2 \lesssim
\alpha \lesssim -1.5$ (e.g., Whitmore \& Schweizer 1995; Elmegreen \&
Efremov 1997; Miller et al. 1997; Whitmore et al. 1999; Whitmore et
al. 2002; Bik et al. 2003; de Grijs et al. 2003; Hunter et al. 2003;
Lee \& Lee 2005; see also Elmegreen 2002). Since the spectral index,
$\alpha$, of the observed CLFs resembles the slope of the
high-mass regime of the (lognormal) old GC mass spectrum ($\alpha \sim
-2$; McLaughlin 1994), this observational evidence has led to the
popular theoretical prediction that not only a power law, but {\it
any} initial CLF (and CMF) will be rapidly transformed into a
lognormal distribution because of (i) stellar evolutionary fading of
the lowest-luminosity (and therefore lowest-mass, for a given age)
objects to below the detection limit; and (ii) disruption of the
low-mass clusters due to both interactions with the gravitational
field of the host galaxy, and internal two-body relaxation effects
leading to enhanced cluster evaporation (e.g., Elmegreen \& Efremov
1997; Gnedin \& Ostriker 1997; Ostriker \& Gnedin 1997; Fall \& Zhang
2001; Prieto \& Gnedin 2007).

However, because of observational selection effects it is often
impossible to probe the CLFs of YMC systems to the depth required to
fully reveal any useful evolutionary signatures. As such, the young
populous cluster systems in the Magellanic Clouds are as yet the best
available calibrators for the canonical young CLFs that form the basis
of most theoretical attempts to explain the evolution of the CLF and
CMF.\footnote{We point out, however, that with the latest {\sl
HST}/Advanced Camera for Surveys observations of the Antennae
interacting galaxies, we are now finally getting to the point where
the young (power-law) CLF shape appears to be confirmed independently
down to sufficient photometric depths and for larger samples
containing more massive clusters than in the Magellanic Clouds
(B. Rothberg, priv. comm.).} It is therefore of paramount importance
to understand the Magellanic Cloud cluster systems in detail.

\section{Early cluster evolution in the Magellanic Clouds}

The early evolution of the star cluster population in the Small
Magellanic Cloud (SMC) has been the subject of considerable recent
interest (e.g., Rafelski \& Zaritsky 2005; Chandar, Fall \& Whitmore
2006; Chiosi et al. 2006; Gieles, Lamers \& Portegies Zwart 2007; de
Grijs \& Goodwin 2008). The key issue of contention is whether the
SMC's star cluster system has been subject to the significant early
cluster disruption processes observed in `normal', interacting and
starburst galaxies, commonly referred to as `infant mortality'. Chandar
et al. (2006) argue that the SMC has been losing up to 90\% of its
star clusters per decade of age, at least for ages from $\sim 10^7$ up
to $\sim 10^9$ yr, whereas Gieles et al. (2007) conclude that there is
no such evidence for a rapid decline in the cluster population, and
that the decreasing number of clusters with increasing age is simply
caused by evolutionary fading of their stellar populations in a
magnitude-limited cluster sample.

\begin{figure}[h]
\begin{center}
 \includegraphics[width=\columnwidth]{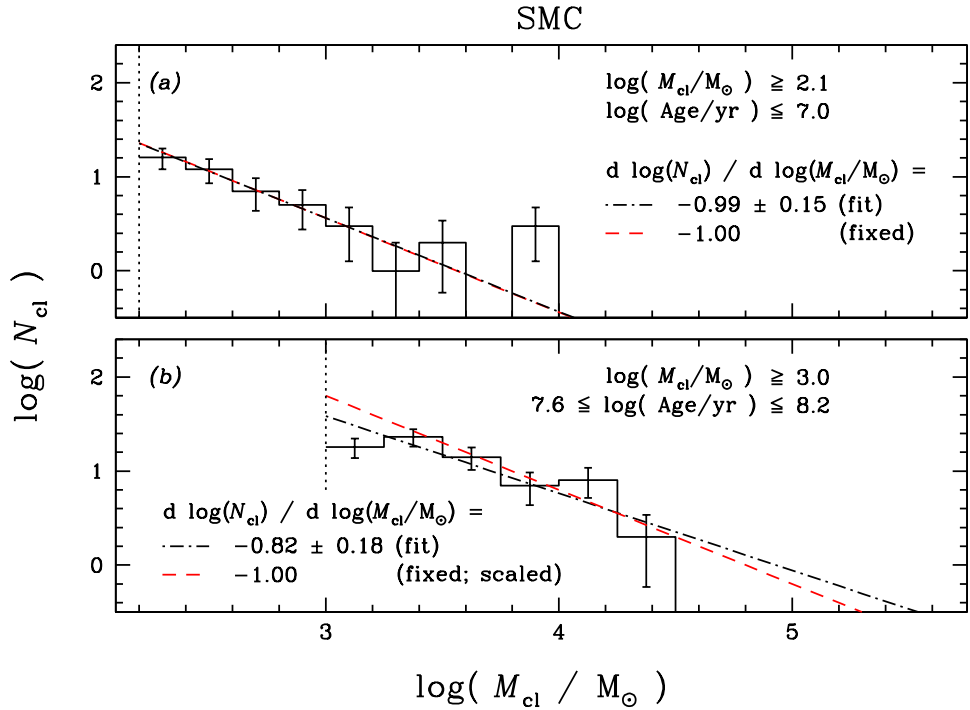} 
\vspace*{-4cm}
 \caption{CMFs for statistically complete SMC cluster subsamples. Age
and mass ranges are indicated in the panel legends; the vertical
dotted lines indicate the lower mass (50\% completeness) limits
adopted. Error bars represent simple Poissonian errors, while the
dash-dotted lines represent CMFs of slope $\alpha = 2$, shifted
vertically as described in the text.}
\end{center}
\end{figure}

In de Grijs \& Goodwin (2008) we set out to shed light on this
controversy. We adopted a fresh approach to the problem, using an
independent, homogeneous data set of $UBVR$ imaging observations
(cf. Hunter et al. 2003), from which we obtained the cluster age
distribution in a self-consistent manner. In Fig. 1 we present the
CMFs for two subsets of our SMC cluster sample, selected based on
their age distributions. In all panels of Fig. 1, we have overplotted
CMFs with the canonical slope of $\alpha = -2$ (corresponding to a
slope of $-1$ in units of ${\rm d} \log(M_{\rm cl}/{\rm M}_\odot) /
{\rm d} \log (N_{\rm cl})$, used in these panels).

The rationale for adopting as our youngest subsample (Fig. 1a) all
clusters with ages $\le 10$ Myr is that at these young ages, the vast
majority of the star clusters present will still be detectable, even
in the presence of early gas expulsion (e.g., Goodwin \& Bastian 2006)
-- as long as they are optically conspicuous. Fig. 1b includes our
sample clusters with ages from 40 Myr to 160 Myr. While the upper age
limit ensures the full inclusion of the clusters affected by the onset
of the AGB stage, its exact value is rather unimportant. The lower age
limit of this subsample is crucial, however. As shown by Goodwin \&
Bastian (2006), most dissolving clusters will have dispersed by an age
of $\sim 30$ Myr, while the surviving clusters will have returned to
an equilibrium state by $\sim 40$ Myr, when some of the early
expansion will have been reversed, depending on the effective
star-formation efficiency. This latter age is therefore a good lower
boundary to assess the surviving star cluster population.

We explicitly exclude any star clusters aged between 10 and 40 Myr
from our analysis. In this age range, it is likely that dissolving
star clusters that will not survive beyond about 30--40 Myr might
still be detectable and therefore possibly contaminate our sample. In
addition, this is the age range in which early gas expulsion causes
rapid cluster expansion, before settling back into equilibrium at
smaller radii; because of the expanded nature of at least part of the
cluster sample, we might not be able to detect some of the
lower-luminosity (and hence lower-mass) clusters that would again show
up beyond an age of $\sim 40$ Myr.

\begin{figure}[h]
\begin{center}
 \includegraphics[width=0.8\columnwidth]{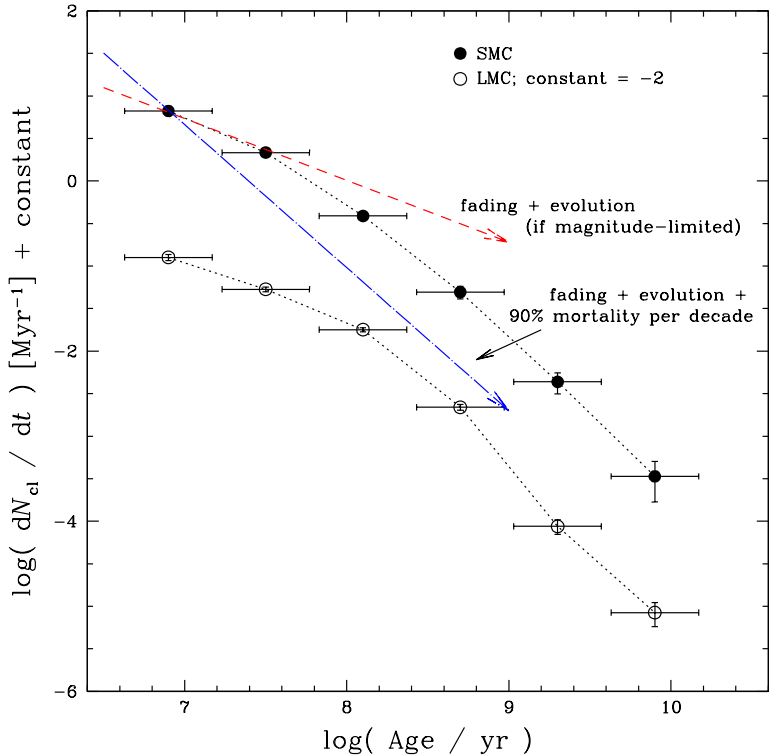} 
 \caption{Age distribution of the full magnitude-limited SMC and LMC
cluster samples in units of cluster numbers per Myr. The LMC sample
has been shifted vertically by a constant offset for reasons of
clarity. The vertical error bars are simple Poissonian errors; the
horizontal error bars indicate the age range used for the generation
of these data points. The dashed arrow shows the expected effects due
to evolutionary fading of a magnitude-limited cluster sample; the
dash-dotted arrow represents the combined effects of a fading cluster
population and 90\% cluster disruption per decade in $\log( \mbox{Age
yr}^{-1} )$.}
\end{center}
\end{figure}

In the simplest case, in which the cluster formation rate has remained
roughly constant throughout the SMC's evolution (see, e.g., Boutloukos
\& Lamers 2003, their fig. 10; see also Gieles et al. 2007), the
number of clusters would simply scale with the age range covered. In
Fig. 1b we show the canonical $\alpha=-2$ CMF scaled from the
best-fitting locus in Fig. 1a by the difference in (linear) age range
between the panels (see de Grijs \& Goodwin 2008 for details). The
scaled canonical CMF in Fig. 1b is an almost perfect fit to the
observational CMF (irrespective of the mass binning employed). This
implies that the SMC cluster system has not been affected by any
significant amount of cluster infant mortality for cluster masses
greater than a few $\times 10^3$ M$_\odot$. Based on a detailed
assessment of the uncertainties in both the CMFs and the age range
covered by our youngest subsample, we can limit the extent of infant
mortality between the youngest and the intermediate age range to a
maximum of $\lesssim 30$\% ($1\sigma$). We rule out a $\sim 90$\%
(infant) mortality rate per decade of age at a $>6 \sigma$ level. This
result is in excellent agreement with that of Gieles et al. (2007).

Moreover, Gieles et al. (2007) derived that for a magnitude-limited
sample the decline in the number of observed clusters per unit age
range, ${\rm d}N_{\rm cl} / {\rm d}t$, as a function of age is
graphically described by a slope of $-0.72$. In Fig. 2, we show the
expected effects of evolutionary fading of a magnitude-limited cluster
population as the dashed arrow. It is immediately clear that, for the
SMC cluster population analysed in de Grijs \& Goodwin (2008), the
decline in the age distribution up to $\log( \mbox{Age yr}^{-1} )
\simeq 7.8$ can indeed be entirely attributed to evolutionary
fading. The expected effects of evolutionary fading combined with a
90\% disruption rate are shown as the dash-dotted arrow in Fig. 2. The
arrow clearly does not fit the observed age distribution, if we
require it to pass through our youngest data point. We note, however,
that the slope of this latter arrow is very similar to that of the age
distribution of the full SMC sample for ages in excess of a few
$\times 10^8$ yr, when secular disruption is likely to take over.

\section{Cluster disruption in the LMC}

To derive the CMF of mass-limited LMC cluster subsamples, de Grijs \&
Anders (2006) re-analysed the {\sl UBVR} broad-band data of Hunter et
al. (2003). They derived that the timescale on which a $10^4 M_\odot$
cluster is expected to disrupt is $\log(t_4^{\rm dis}{\rm
yr}^{-1})=9.9\pm0.1$, as recently confirmed by Parmentier \& de Grijs
(2008) based on a detailed comparison with numerical
simulations.\footnote{In fact, Parmentier \& de Grijs (2008) concluded
that the data do not allow us to constrain the characteristic cluster
disruption timescale for a $10^4$ M$_\odot$ cluster to better than
$9.0 \lesssim \log(t_4^{\rm dis}{\rm yr}^{-1}) \lesssim 9.9$.} Such a
long cluster disruption timescale results from the low-density
environment of the Magellanic Clouds. It guarantees that the observed
cluster mass distributions have not yet been altered significantly by
secular dynamical evolution, i.e. clusters already affected by ongoing
disruption have faded to below the completeness limit (see de Grijs \&
Anders 2006, their fig. 8). As a result, the observed mass
distributions are the {\it initial} distributions.

We will investigate the possible effects of infant mortality among the
LMC cluster population in detail in a forthcoming paper (Goodwin et
al., in prep.). However, a first glance at the LMC cluster
population's age distribution in Fig. 2 indicates that the number of
clusters populating the first $\sim 10^8$ yr can likely be fully
explained by simple evolutionary fading of a magnitude-limited cluster
sample, as we also concluded for the SMC, without the need to invoke
infant mortality for masses $M_{\rm cl} \gtrsim 10^3$ M$_\odot$.

\begin{figure}[t]
\begin{center}
 \includegraphics[width=0.6\columnwidth,angle=-90]{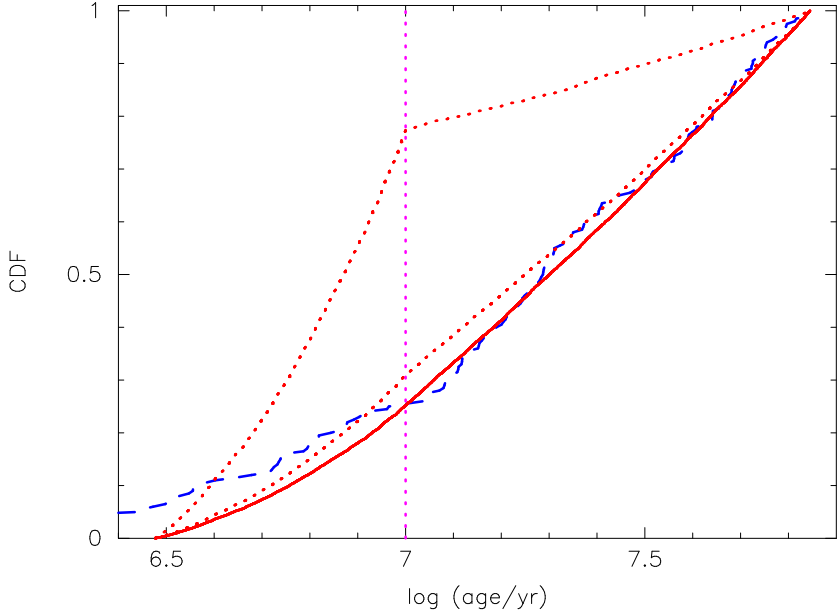} 
 \caption{Cumulative distribution function (CDF) of the LMC cluster
 age distribution. The dashed line represents the data, smoothed by
 the relevant age-dependent uncertainties; the solid line is the fit
 for ages $> 10^7$ yr, assuming no disruption. The dotted lines are
 the expected CDFs for 90\% (top) and 25\% (bottom) mortality per
 decade in age.}
   \label{fig1}
\end{center}
\end{figure}

In Fig. 3 we display the LMC cluster population's cumulative
distribution function (CDF) as a function of age (dashed line). This
CDF has been smoothed by the uncertainties in the cluster ages
determined for each cluster individually by de Grijs \& Anders
(2006). We compare the observed CDF to that of a model with a constant
cluster-formation rate with and ICMF index of $\alpha=-2$. These
clusters are evolved according to the Lamers et al. (2007) disruption
scenario without invoking infant mortality, assuming standard stellar
evolution. Since cluster ages $\lesssim 10$ Myr are highly uncertain,
only the fit beyond 10 Myr should be considered (solid line).

The red dotted lines illustrate what would be expected for 25\%
(bottom) and 90\% (top) sudden infant mortality at an age of 10 Myr
(slightly extended mortality would result in a more smoothed
appearance of these lines). The best-fit model at 10 Myr and
afterwards suggests, for a constant cluster-formation rate, $< 10$\%
mass-independent infant mortality.

We are currently investigating a number of possible caveats that may
be associated with this preliminary analysis. In particular, could the
apparent absence of infant mortality be hidden by our assumption of a
constant cluster-formation rate? Alternatively (or additionally),
could mass-dependent infant mortality be at work in the LMC (and
possibly the SMC)?

\begin{acknowledgments}
We are grateful to Peter Anders and Genevi\`eve Parmentier for
essential contributions.
\end{acknowledgments}

\end{document}